\documentclass[osajnl,twocolumn,showpacs,superscriptaddress,10pt]{revtex4-1} 
\usepackage{amsmath,amssymb,graphicx}
\begin{document}

\title{Simultaneous VUV and XUV pulse generation and characterization for attosecond pump probe experiments }

\author{D. Fabris}\email{Corresponding author: d.fabris11@imperial.ac.uk}
\author{W. A. Okell}
\author{D. Walke}
\author{T. Witting}
\author{J. P. Marangos}
\author{J. W. G. Tisch}
\affiliation{Blackett Laboratory, Imperial College, London SW7 2AZ, UK}
\begin{abstract}
We report the generation and characterization of isolated attosecond XUV and VUV pulses generated simultaneously via HHG driven by few-cycle pulses using  an in-line dual gas target system. One gas jet target was operated with Kr gas that optimized HHG in the 15-25 eV photon energy range (VUV), whilst the second gas jet target was operated in Ne gas to optimize the high harmonic generation around 90 eV (XUV). Appropriate filters were used to isolate the required spectral components to synthesize isolated pulses. Sn and In filters were used for the VUV region while a Zr filter was used for the XUV. We characterized both the XUV and VUV pulses independently using the attosecond streaking technique and the LSGPA retrieval algorithm obtaining a 1.7$\pm0.2$ fs pulse using the In filter and a 616$\pm50$ as pulse using Sn, while preserving a $266\pm10$ as isolated XUV pulse.
\end{abstract}


\maketitle 

\section{Introduction}
A significant part of future development in attosecond science \cite{krausz_attosecond_2009} depends on attosecond light sources in new spectral regions to be applied in novel pump probe experiments \cite{wang_attosecond_2009}. The production of isolated attosecond pulses (IAP) in the XUV spectral region (30-150 eV) is nowadays a robust process. It requires gating the high harmonics (HH) emission produced with an intense laser using different techniques. Most common are polarization gating \cite{mashiko_double_2008}, ionization gating \cite{ferrari_high-energy_2010} and amplitude gating \cite{witting_sub-4-fs_2012}. However the yield of photons of such pulses is typically too low for them to be used in an attosecond-pump attosecond probe experiment. Therefore the driving IR field is usually employed with the XUV pulse in a femtosecond-attosecond pump-probe scheme.
Using the IR field, however, lowers the time resolution of the experiment, moreover it is not always the best choice to study attosecond dynamics as many target molecules and atoms of interest show strong perturbation in a IR field of sufficient strength to induce photoionisation. 
 
 The results presented in this paper focus on the VUV spectral region (10-25 eV). The use of isolated pulses in this spectral region is advantageous for use in the pump step of a pump-probe measurement. Compared to HH generated in higher energy spectral regions  ($>$50 eV) the HH generation in the 10-25 eV region is expected to provide a far higher photon yield and have a much higher interaction cross-section (e.g. for photoionisaton) with a typical atomic or molecular species\cite{kameta_photoabsorption_2002}. 
 
 For these reasons there is a great deal of interest in developing reliable higher power attosecond sources both at higher \cite{popmintchev_bright_2012} or lower \cite{mashiko_tunable_2010,beutler_generation_2011,graf_intense_2008,feng_generation_2009} energies with respect to the standard XUV region.
 
 The metallic filters used to select a portion of the HH spectrum generated with a few-cycle IR pulse are indium (central energy 15 eV) and tin (central energy 20 eV) \cite{cxro}. Using a collinear target geometry has been proven to be a promising approach \cite{bothschafter_collinear_2010}. We have used collinear pulse gas targets to simultaneously generate an isolated XUV pulse ($\tau_{XUV}=266\pm10$ as) together with VUV pulses ($\tau_{\text{15eV}}=1.7\pm0.2$ fs, $\tau_{\text{20eV}}=616\pm50$ as). The pulses were characterised using IR-field dressed photoionisation and the FROG-CRAB method [14] for both the XUV-IR and VUV-IR configurations.

\section{Experiment}
\begin{figure}[htbp]
\centerline{\includegraphics[width=.9\columnwidth]{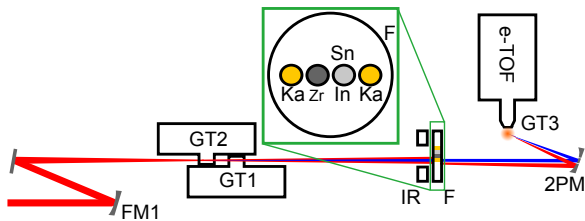}}
\caption{Schematics of experimental setup. FM1,FM2: Focusing mirrors. GT1, GT2: pulsed gas targets. IR: iris. F: filters, GT3: effusive gas target, 2PM: MoSi 2 part mirror, e-TOF: electron time of flight spectrometer.}\label{fig:setup}
\end{figure}

The experiment reported in this paper was performed using the Imperial College attosecond beamline \cite{frank_invited_2012}. A commercial chirped pulse amplification laser system (Femtolaser GmbH, Femtopower HE CEP) provided pulses at 1 kHz, 28 fs duration with up to 2.5 mJ per pulse. The CPA pulses were compressed to 3.5 fs duration and pulse energy of up to 800 $\mu$J in a differentially pumped hollow core fibre (HCF)\cite{robinson_generation_2006} in order to implement amplitude gating  in the HHG process. The carrier-envelope phase (CEP) of the pulses was locked after the fibre output with a feedback system on the oscillator pump \cite{okell_carrier-envelope_2013}. Laser pulses of 400 $\mu$J were focused with a 70 cm focusing mirror (FM1) onto two in line pulsed gas targets as shown in Figure \ref{fig:setup}.

The pulsed gas target GT1 was Ne (at a gas density of $\approx10^{17}\text{cm}^{-3}$) for the production of the XUV radiation. The second pulsed gas target GT2 was Kr (at a gas density of $\approx10^{16}\text{cm}^{-3}$), whose lower ionization potential results in a more efficient production of VUV radiation. The position of the two targets with respect to the laser focus and with respect to each other was optimized for phase-matching by performing position scans of the jets along the propagation axis. To filter the HH radiation a set of metallic filters was mounted on a stage for convenient interchange. Kapton (7.5 $\mu$m thickness) was used to reject the HH radiation providing the IR pulse for the pump probe experiment. A Zr  filter (200 nm thickness) rejected the IR and lower harmonics filtering the XUV pulse. An In filter (200 nm thickness) selected a VUV pulse centered at 15 eV while Sn (200 nm thickness) filtered a VUV pulse centered at 20 eV.   The mounting geometry allowed us to select all pairs of pump-probe pulses in combination with the filters. The filter mount was positioned so that the beam was intercepted by two adjacent filters (e.g. Kapton and In to streak the VUV at 15 eV),  leading to side-by-side co-propagating portions of the beam that have been differently filtered.  

The streaking setup \cite{frank_invited_2012} included a MoSi multilayer two part mirror mounted on a piezo stage that selected a 8.6 eV bandwidth around 92 eV for the XUV pulse and provided the time delay between the pulses. The two time delayed pulses were focused onto the effusive gas target in front of the electron time of flight spectrometer (e-TOF). Photoelectron spectra were recorded as a function of time delay between the XUV/VUV pulses and the IR pulse. 

Though we have not yet made an absolute measurement of photon fluxes of the pulses, based on known photoionisation cross-sections \cite{kennedy_photoionization_1972}, filter trasmission and the collection efficiency of our electron spectrometer, and assuming the MoSi multilayer mirror to have a reflectivity in the VUV of few \% (2\% at 30 eV \cite{cxro}), we estimate that we generate $\approx$ 10 nJ of VUV per pulse.

\section{Results}
\subsection{XUV-VUV compatibility}
\begin{figure}[htbp]
\centerline{\includegraphics[width=.9\columnwidth]{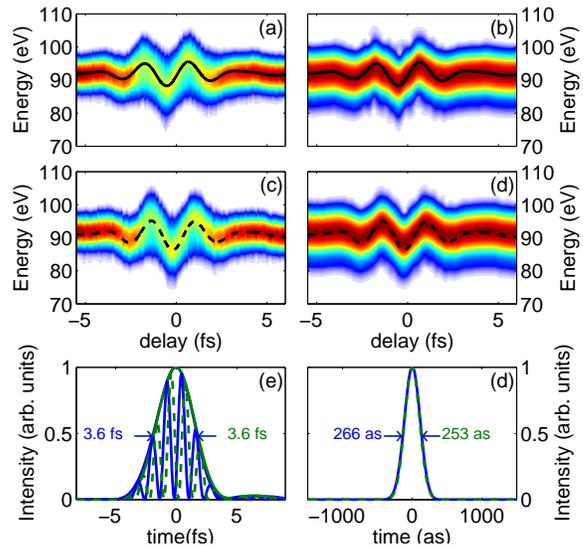}}
\caption{Effect of second gas target. XUV-IR measured and retrieved traces with gas target off (a-b) and on (c-d). (e) IR pulses, with gas target off (blue line) and on (green dashed line). (d) XUV retrieved pulses from (a) (blue line) and (c) (green dashed line).}\label{fig:streak_comparison}
\end{figure}

The first result achieved is the proof that the inline geometry is compatible with simulataneous attosecond XUV pulse generation and VUV pulse generation. For this reason XUV-IR streaking traces in Ne were recorded both with the gas target for the VUV pulse (which we subsequently refer to as GT2) on and off. The results are shown in Figure  \ref{fig:streak_comparison}. Panels (a) and (b) show the measured and retrieved XUV-IR traces with GT2 off, while in panel (c) and (d) the same situation is illustrated for GT2 on. Panel (e) shows the retrieved IR pulses, showing that the effect of the VUV gas target is a CEP shift, while its duration is not altered. Panel (d) shows the retrieved XUV pulses. The two traces have been analyzed with exactly the same smoothing parameters. The algorithm implemented for the retrieval is LSGPA \cite{gagnon_accurate_2008} and it has been applied in the same way to both traces providing a time duration of $266\pm10$ as and $253\pm10$. The stated errors include both the retrieval and experimental uncertainties. A difference in time duration of $\approx$13 attoseconds is therefore within the errors showing that the XUV pulse was not strongly perturbed by the presence of the VUV gas target maintaining a pulse duration sufficiently short for attosecond pump probe experiments.

\subsection{VUV pulse at 15 eV}
\begin{figure}[htbp]
\centerline{\includegraphics[width=.9\columnwidth]{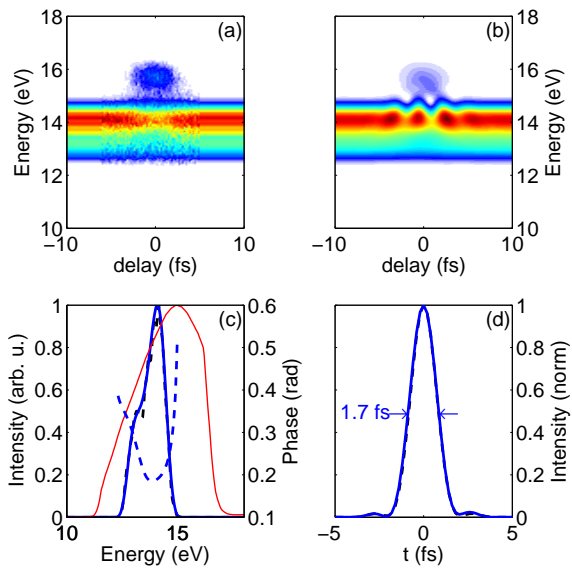}}
\caption{Results using an indium filter. (a) Measured trace. (b) Retrieved trace. (c) Spectral intensity, retrieved (blue), measured (black dashed), phase (dashed blue), transmission of In filter (red). (d) Temporal intenisty, retrieved (blue) FTL (black dashed).}\label{fig:In}
\end{figure}

A 200 nm thick foil of indium provides a transmission of about 25\% at 15 eV, with a bandwidth of $\approx$ 3 eV, corresponding to a Fourier Transform limit (FTL) of 820 as. Theoretical studies \cite{henkel_prediction_2013} showed such a filter can be used to select an isolated attosecond pulse. The results obtained with an indium filter are shown in Figure \ref{fig:In}.

The traces were obtained using xenon as gas target to allow single photon ionization by the VUV pulse. Panels (a) and (b) are the measured and retrieved trace respectively. The agreement between measured and retrieved data is good with a FROG error of 1.7\% on a grid 160$\times$1024.  Panel (c) shows the retrieved intensity and phase in the spectral domain with the blue continuous and dashed line respectively. The dashed black line is the measured spectrum and the red line is the normalized In transmission. Panel (d) shows with the blue curve the retrieved pulse, while the black-dashed one is the FTL from the IR-free spectra measured at the lowest delays at the beginning of the trace. 

The retrieved pulse duration is $1.7\pm0.2$ fs, much longer than the FTL of the spectral bandwidth provided by the In filter. This is to be expected since we are implementing amplitude gating only, hence the HH spectrum at low energy is rather discrete. In this case only one harmonic(9th at 14 eV) was selected, resulting in a well-isolated pulse without noticeable satellites. With a 3.5 fs IR pulse, assuming the pulse duration of each harmonic scales as $1/\sqrt{q}$ where $q$ is the harmonic order, the expected VUV pulse duration is $\approx$1.2 fs. Some features of the retrieved trace are not present in the experimental data. This can be attributed to the energy resolution of the time-of-flight spectrometer (TOF) not being optimized for such low energies (actual photoelectron energy measured between $\approx$1-4 eV).

\subsection{VUV pulse at 21 eV}
\begin{figure}[htbp]
\centerline{\includegraphics[width=.9\columnwidth]{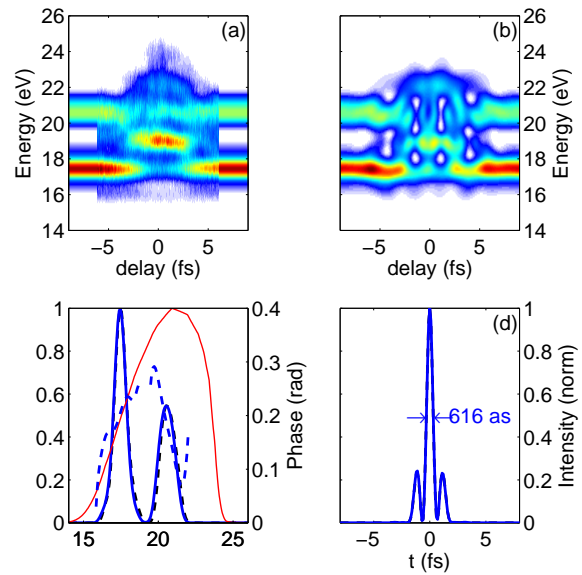}}
\caption{Results using a tin filter. Plots as in figure \ref{fig:In}.}\label{fig:Sn}
\end{figure}

The 200 nm tin foil provides 22\%  transmission at 21 eV, with a bandwidth of $\approx$ 5 eV, corresponding to a FTL pulse of 500 as. The results obtained with a tin filter are shown in Figure \ref{fig:Sn}.

The target was Xe in this case as well. The plots and color code of Figure \ref{fig:Sn} are the same as in Figure \ref{fig:In}. The agreement between measured and retrieved traces is good with a FROG error of 3\% on a grid 160$\times$512. In this case the retrieved pulse has a duration of $616\pm50$ as. The presence of two harmonics in the spectral window provided by tin leads to the presence of satellite pulses in the time domain. However they are a factor of 5 lower in intensity with respect the main pulse. Again the absence of some features in the experimental data with respect to the retrieval can be attributed to the spectrometer resolution, even though in this case, given the photoelectron energies measured are higher ($\approx$5-9 eV), a hint of the retrieved features can be observed in the experimental trace. 


\section{Conclusions}
The compatibility of VUV and XUV IAP in a collinear gas target geometry has been proven by measuring an IAP in the XUV both with the gas for the VUV generation present and absent with no appreciable change in the measured XUV pulse duration. 

Indium has been used to filter the HH spectrum, providing a $1.7\pm0.2$ fs pulse. This pulse duration is similar to the value expected from a scaling law for the pulse duration following $1/\sqrt{q}$ (1.2 fs).

Tin has been used to filter a pulse at 20 eV. The measured pulse duration is $616\pm50$ as.

This is the first time, to best of our knowledge, that  IAP pulses in the 15-25 eV spectral region have been generated and characterized together with an IAP in the XUV. The results demonstrate the feasibility of a HH based XUV-VUV attosecond-attosecond pump-probe experiments at these photon energies.

\section*{Acknowledgments}
This work was financially supported by EPSRC through grants EP/I032517/1 and EP/F034601/1. We acknowledge technical support from Andrew Gregory and Peter Ruthven.

\end{document}